\documentclass[11pt]{article}

\usepackage{amssymb}
\usepackage{amsmath}
\usepackage{graphicx}
\usepackage{hep}
\usepackage[labelfont=bf]{caption}

\newcommand{\hzero}{\ensuremath{\PHiggslightzero}} 
\newcommand{\Hzero}{\ensuremath{\PHiggsheavyzero}} 
\newcommand{\Azero}{\ensuremath{\PHiggspszero}} 
\newcommand{\Hpm}{\ensuremath{\PHiggspm}} 
\newcommand{\eeHHh}{\ensuremath{\HepProcess{\APelectron\,\Pelectron \HepTo \PHiggsplus \,\PHiggsminus\,\hzero}}}
\newcommand{\eeHHHzero}{\ensuremath{\HepProcess{\APelectron\,\Pelectron \HepTo \PHiggsplus\,\PHiggsminus\,\Hzero}}}

\newcommand{\eehhA}{\ensuremath{\HepProcess{\APelectron\,\Pelectron \HepTo \hzero \,\hzero\,\Azero}}}

\newcommand{\eeHzeroAh}{\ensuremath{\HepProcess{\APelectron\,\Pelectron \HepTo \Hzero \,\Azero\,\hzero}}}
\newcommand{\CP}{\ensuremath{\mathcal{C}\mathcal{P}}}

\newcommand{\sbt}{\sin\beta}

\newcommand{\cbt}{\cos\beta}

\newcommand{\stbt}{\sin2\beta}
\newcommand{\ctbt}{\cos2\beta}

\newcommand{\sbma}{\sin (\beta-\alpha)}
\newcommand{\cbma}{\cos (\beta-\alpha)}

\newcommand{\mw}{M_W}

\newcommand{\mh}{M_{h^0}}
\newcommand{\mH}{M_{H^0}}
\newcommand{\mHp}{M_{H^\pm}}
\newcommand{\mA}{M_{A^0}}

\newcommand{\mHs}{M^2_{H^0}}
\newcommand{\mHps}{M^2_{H^\pm}}
\newcommand{\mAs}{M^2_{A^0}}

\newcommand\pubblock{\rightline{\begin{tabular}{l} \pubnumber\\
     \pubdate\\  \end{tabular}}}
\newcommand\pubnumber{UB-ECM-PF-07/22}
\newcommand\pubdate{July 2007 }

\def\beq{\begin{eqnarray}}    
\def\eeq{\end{eqnarray}}      

\newcommand{\mysection}[1]{\section{#1}
\renewcommand{\theequation}{\thesection.\arabic{equation}}
\setcounter{equation}{0}}

\begin{document}
\pubblock


 \hyphenation{cos-mo-lo-gi-cal
sig-ni-fi-cant}




\begin{center}
{\large \textsc{Triple Higgs boson production in the Linear
Collider}} \vskip 2mm

 \vskip 8mm

\textbf{Giancarlo Ferrera}$^{a}$, \textbf{Jaume Guasch}$^{b,c}$,
\textbf{David L\'opez-Val}$^{a}$, \textbf{Joan Sol\`{a}}$^{a,c}$

\vskip0.5cm

$^{a}$ High Energy Physics Group, Dept. ECM, Univ. de Barcelona\\
Av. Diagonal 647, E-08028 Barcelona, Catalonia, Spain

$^{b}$ Gravitation and Cosmology Group,  Dept. FF, Univ. de
Barcelona\\  Av. Diagonal 647, E-08028
    Barcelona, Catalonia, Spain \\

$^{c}$ Institut de Ci\`{e}ncies del Cosmos, UB, Barcelona.

E-mails: ferrera@ecm.ub.es, jaume.guasch@ub.edu, dlopez@ecm.ub.es,
sola@ifae.es. \vskip2mm

\end{center}
\vskip 15mm

\begin{quotation}
\noindent {\large\it \underline{Abstract}}.\ \ Triple Higgs boson
production (3H) may provide essential information to reconstruct the
Higgs potential. We consider 3H-production in the International
Linear Collider (ILC) both in the Minimal Supersymmetric Standard
Model (MSSM) and in the general Two-Higgs-doublet Model (2HDM). We
compute the total cross-section for the various 3H final states,
such as $\PHiggsplus\,\PHiggsminus\,\hzero$,\
$\Hzero\,\Azero\,\hzero$, etc. and compare with the more traditional
double Higgs (2H) boson production processes. While the
cross-sections for the 2H final states lie within the same order of
magnitude in both the MSSM and 2HDM, we find that for the 3H states
the maximum 2HDM cross-sections, being of order $0.1\,$pb, are much
larger than the MSSM ones which, in most cases, are of order
$10^{-6}\,$ pb or less. Actually, the 3H processes could be the
dominant mechanism for Higgs boson production in the 2HDM.
Ultimately, the origin of the remarkable enhancement of the 3H
channels in the 2HDM case (for both type I and type II models)
originates in the structure of the trilinear Higgs boson couplings.
The extremely clean environment of the ILC  should allow a
relatively comfortable tagging of the three Higgs boson events. In
view of the fact that the MSSM contribution is negligible, these
events should manifest themselves mainly in the form of $6$
heavy-quark jet final states. Some of these signatures could be
spectacular, and in case of being detected  would constitute strong
evidence of an extended Higgs sector of non-supersymmetric origin.

\end{quotation}
\vskip 8mm



\vskip 6mm

 \noindent \mysection{Introduction}
 \label{Introduction}

There is no doubt that the Higgs sector is the most paradigmatic one
in the structure of any modern quantum field theory (QFT) aiming at
a good phenomenological description of electroweak interactions in
particle physics. The main reason for this is twofold: i) the Higgs
mechanism is the only known consistent quantum field theoretical
procedure to generate masses for all the elementary particles; ii)
we have found no Higgs boson yet -- not even the single one
predicted by the successful standard model (SM) of the strong and
electroweak interactions--, and therefore we don't know if Higgs
bosons exist at all or if, on the contrary, there are extensions of
the SM containing a richer spectrum of Higgs boson particles, some
of them electrically charged and some of them electrically neutral.
Let us note that if failure of point ii) would persist for long,
especially after the LHC and the future linear colliders ILC and/or
CLIC had already amply swept their maximum energy ranges and
luminosities, we could find ourselves in a sort of
\textit{cul-de-sac} because this would also mean that we would not
have substantiated point i) either, which is tantamount to say that
we would have not found any experimental evidence of the most
powerful theoretical mechanism known up to date for building
renormalizable models of particle interactions. It is therefore a
momentous task to search for Higgs bosons and unveil their ultimate
nature.

Surely a linear $e^+e^-$ collider will be instrumental to accomplish
this aim because it is the cleanest high-precision machine we can
think of for studying particle interactions. No doubt, if Higgs
bosons are around, the linear collider will help either to discover
them or to identify the precise nature of the Higgs particle(s)
previously uncovered at the LHC. In particular, once a neutral Higgs
boson has been identified, we would like to know if it is the
neutral SM Higgs boson, or if it belongs to some supersymmetric
(SUSY) extension of the SM, or if on the contrary it has nothing at
all to do with SUSY. If, alternatively, the identified Higgs boson
is charged we would like to know to which extension of the SM it can
be ascribed. A particularly well-motivated possibility along these
lines is the Minimal Supersymmetric Standard Model
(MSSM)\,\cite{susy}. But another, simpler, one is just the general
(unconstrained) Two-Higgs-Doublet Model (2HDM)\,\cite{hunter}.

Double Higgs boson (2H) production in a linear collider has been
investigated in great detail in the literature, although mainly in
the MSSM\,\cite{Djouadi:1992pu,pairmssm,Feng:1996xv}. Such process
cannot proceed in the SM at the tree-level, so we know that if we
would detect a sizeable rate of 2H final states in a
$\APelectron\Pelectron$ collider it would be an unmistakable sign of
physics beyond the SM. However, a tree-level analysis of these
pairwise-produced unconventional Higgs bosons is most likely
insufficient to unravel their true nature. Therefore, a dedicated
work on radiative correction calculations has been undertaken. A
rich literature exists indeed on the one-loop calculation of the
cross-sections for the two-particle final states
\begin{eqnarray}
  \APelectron\Pelectron \to 2\PHiggs\,\ \ \ \ \ (2\PHiggs \equiv \hzero\,\Azero; \Hzero\,\Azero;
\PHiggsplus\PHiggsminus)\,, \label{2H}
\end{eqnarray}
essentially in the MSSM case\,\footnote{See
\cite{mssmloop,Djouadi:1999gv,Fawzy02}, and also the extensive
report \cite{Weiglein:2004hn} and references therein.}. Similarly,
the two-body final states $\APelectron\Pelectron \to \PZ
\PHiggslight$ and $\APelectron\Pelectron \to \PHiggsps\PHiggslight$
(with $\PHiggslight = \hzero \Hzero$) are long known to be
complementary to each other in the MSSM\,\cite{Djouadi:1992pu}.
There are also studies considering radiative corrections to charged
Higgs production in $\APelectron\Pelectron$ collisions within the
2HDM\,\cite{ghk}, and double and multiple Higgs production at the
LHC\,\cite{inlhc}, but to our knowledge a complete analysis of the
processes (\ref{2H}) in the general 2HDM is lacking\,\cite{paper2}.

In another vein, triple Higgs boson (3H) production may open new
vistas in our desperate hunting for the mass generation mechanism.
These processes can be very important because they carry essential
information to reconstruct the Higgs boson potential and thus of the
Higgs mechanism itself. The Higgs potential of any renormalizable
QFT may contain in general mass terms, trilinear Higgs boson
self-interactions and quartic self-interactions. For instance, the
trilinear coupling $\PHiggsheavy\PHiggsheavy\PHiggsheavy$ has been
investigated phenomenologically in TeV-class linear colliders in
Ref. \cite{pairmssm,Djouadi:1999gv,Fawzy02} through the double-Higgs
strahlung process $\APelectron\Pelectron \to
\PHiggsheavy\PHiggsheavy \PZ$ or the $\PW\PW$ double-Higgs fusion
mechanism $\APelectron\Pelectron \to \PHiggsplus\PHiggsminus
\Pnue\APnue$. These processes involve vertices like
$\PZ\PZ\PHiggsheavy$, $\PW\PW\PHiggsheavy$,
$\PZ\PZ\PHiggsheavy\PHiggsheavy$, $\PW\PW\PHiggsheavy\PHiggsheavy$
and $\PHiggsheavy\PHiggsheavy\PHiggsheavy$, and are possible both in
the SM and in extensions of the SM, like the MSSM and the general
2HDM. Unfortunately the cross-section turns out to be rather small
(of order of a fb at most) both in the SM and in the MSSM
\,\cite{Djouadi:1999gv}. Even worse is the situation with the triple
Higgs boson production in the MSSM, unless in some specific
configuration of the parameter space with resonant enhancement of
the signal, see Section \ref{sect:numerical}. Out of the resonance,
the typical cross-sections are of order of $0.01\ $fb or
less\,\cite{Djouadi:1999gv}. In the previous reference it has been
shown that if the double and triple Higgs production cross sections
would yield sufficiently high signal rates, the system of couplings
and corresponding double/triple Higgs production cross sections
could be solved for all trilinear Higgs self-couplings up to
discrete ambiguities, by using these processes. However, in practice
the cross sections are too small to be all measurable.

In this letter we wish to study the trilinear coupling
$\PHiggsheavy\PHiggsheavy\PHiggsheavy$ in the
general 2HDM case by focusing on exclusive triple Higgs boson final
states produced at the ILC. We find that there are scenarios where
the $\PHiggsheavy\PHiggsheavy\PHiggsheavy$ coupling
could actually be identified relatively easily.
This is because in the general 2HDM it can be highly enhanced as
compared to the MSSM case (which is purely gauge). To show the
phenomenological impact of this enhancement, and also to briefly
compare with the MSSM situation, we compute the 3H production
cross-sections for all possible \CP-conserving final states both in
the MSSM and the 2HDM. The seven allowed triple-Higgs boson channels
can be sorted out in three main classes:
\begin{eqnarray}
1)\ \APelectron\Pelectron \to \PHiggsplus\PHiggsminus \PHiggslight
\, ,\ \ \ 2)\ \APelectron\Pelectron \to \PHiggslight \PHiggslight \Azero\,
, \ \ \ 3)\ \APelectron\Pelectron \to \hzero \Hzero \Azero\,,
\ \ \ (\PHiggslight=\hzero,\Hzero,\Azero) \label{3H}
\end{eqnarray}
where in class 2) we understand that the two neutral Higgs bosons
$\PHiggslight$ must be the same, i.e. the allowed final states are
$(\PHiggslight\, \PHiggslight \Azero)
=(\hzero \hzero \Azero)$, $(\Hzero \Hzero \Azero)$ and $(\Azero \Azero
\Azero)$. We show that the 2HDM cross-sections can be several orders
of magnitude larger than the corresponding MSSM ones.
Interestingly enough, the 3H cross-sections can be comparable and even
significantly larger than the 2H cross-sections irrespective of the
latter being computed in the MSSM or in the 2HDM.

\noindent \mysection{General 2HDM: relevant interactions and
restrictions} \label{sect:interaction}

In this section we shall briefly present the interactions and
phenomenological restrictions relevant to our calculation. Let us
recall that the general 2HDM is obtained by canonically extending
the SM Higgs sector with a second  $SU_L(2)$ doublet with weak
hypercharge $Y=1$, so that it contains $4$ complex scalar fields
arranged as follows:
\begin{equation}
\Phi_1=\left(\begin{array}{c} \Phi_1^{+} \\ \Phi_1^0
\end{array} \right)
\ \ \ (Y=+1)\,,\ \ \ \ \ \Phi_2=\left(\begin{array}{c} \Phi_2^{+} \\
\Phi_2^0
\end{array} \right)\ \ \ (Y=+1) \,\,.
\label{eq:H1H2}
\end{equation}
In the supersymmetric case, in order to construct a consistent
superpotential\,\cite{susy} one replaces $\Phi_1$ with the
conjugate ($Y=-1$) $SU_L(2)$ doublet
\begin{equation}\label{Yminus}
H_1= \left(\begin{array}{c} H_1^0 \\
H_1^{-}\end{array} \right)\equiv\epsilon\,\Phi_1^*=\left(\begin{array}{c} \Phi_1^{0*} \\
-\Phi_1^{-}
\end{array} \right)
\ \ \ \ \ \ (Y=-1)\,.
\end{equation}
Here $\epsilon=i\,\sigma_2$. For simplicity we stick to the
canonical form (\ref{eq:H1H2}) because we need not presume SUSY, the
correspondence with the MSSM case being now clear
($\Phi_1=-\epsilon\,H_1^*$).

\noindent In this framework the most general structure of the
$\mathcal{C}\mathcal{P}$-conserving, gauge invariant, renormalizable
Higgs potential preserving the discrete symmetry $\Phi_i\to
(-1)^i\,\Phi_i\ (i=1,2)$, reads\,\cite{hunter}:
\begin{eqnarray}
&&V(\Phi_1,\Phi_2) = \lambda_1\,(\Phi_1^\dagger\,\Phi_1 - v_1^2)^2 +
\lambda_2\,(\Phi_2^\dagger \Phi_2 - v_2^2)^2 +
\lambda_3\,\left[(\Phi_1^\dagger\,\Phi_1 - v_1^2)
+(\Phi_2^\dagger\,\Phi_2 - v_2^2) \right]^2 \nonumber \\
&& +\lambda_4\,\left[(\Phi_1^\dagger \Phi_1)(\Phi^\dagger_2\Phi_2) -
(\Phi^\dagger_1\Phi_2)(\Phi^\dagger_2\Phi_1)\right] +\lambda_5\,
\left[Re(\Phi_1^\dagger \Phi_2) -v_1\,v_2\right]^2 +  \lambda_6\,
\left[Im(\Phi_1^\dagger \Phi_2)
 \right]^2
\label{eq:potential}
\end{eqnarray}
\noindent $\lambda_i \,(i=1,\dots\,6)$ being dimensionless real
parameters. Once the neutral components of the Higgs doublets
acquire non-vanishing VEV's (vacuum expectation values), the
electroweak (EW) symmetry $SU_L(2)\times U_Y(1)$ breaks down to
$U(1)_{em}$, in such a way that we can obtain the physical spectrum
of the Higgs sector upon diagonalization of
Eq.~(\ref{eq:potential}). We are thus left with two
$\mathcal{C}\mathcal{P}$-even  Higgs bosons $\hzero$, $\Hzero$, a
$\mathcal{C}\mathcal{P}$-odd Higgs boson $\Azero$ and a pair of
charged Higgs bosons $\Hpm$. The free parameters in the general 2HDM
are usually chosen to be as follows: the masses of the physical
Higgs particles ($M_{\hzero}, M_{\Hzero}, M_{\Azero}, M_{\Hpm}$);
the ratio of the two VEV's $\braket{H_i^0}\equiv v_i/\sqrt{2}\,
(i=1,2)$ giving masses to the up- and down-like quarks,
\begin{equation}
\tan \beta \equiv \frac{\braket{H_2^0}} {\braket{H_1^0}} \equiv
\frac{v_2}{v_1}\,; \label{tb}
\end{equation}
\noindent the mixing angle $\alpha$ between the two \CP-even states;
and finally the coupling $\lambda_5$, which cannot be absorbed in
the previous quantities and becomes tied to the structure of the
Higgs self-couplings. In total, we have $7$ free parameters, which
indeed correspond to the original $6$ couplings $\lambda_i$ and the
two VEV's $v_1,v_2$ -- the latter being submitted to the physical
constraint $v^2\equiv v_1^2+v_2^2=\,G_F^{-1}/\sqrt{2}$, where $G_F$
is Fermi's constant (equivalently, $v^2= 4\,M_W^2/g^2$, where $M_W$
is the $\PWpm$ mass and $g$ the $SU_L(2)$ gauge coupling constant).
Incidentally, let us note that the essential parameter (\ref{tb})
could be ideally measured in $\APelectron\Pelectron$
colliders\,\cite{Feng:1996xv} e.g. through production and decay of
$\Hpm$ or $\Azero$ since in these cases the rates do not involve the
mixing angle $\alpha$. It is also worth noticing that two of the
$\lambda_i$ parameters can be directly related to physical Higgs
boson masses and to the Fermi constant: $\lambda_4 =
2\,M_{\Hpm}^2/v^2=2\sqrt{2}\,G_F\,M_{\Hpm}^2$ and $\lambda_6 =
2\,M_{\Azero}^2/v^2=2\sqrt{2}\,G_F\,M_{\Azero}^2$. These relations
are valid only at the tree level.

Let us point out that the aforementioned discrete symmetry imposed
on (\ref{eq:potential}), which is only softly violated by the
dimension-two terms, is necessary to ensure the absence of
tree-level flavor changing neutral currents (FCNC). It is well-known
that this discrete symmetry is automatically preserved in the MSSM.
However in the general case it has to be imposed, and then two main
scenarios arise\,\cite{hunter}: 1) In the so-called type I 2HDM one
Higgs doublet ($\Phi_2$) couples to all of the SM fermions, whereas
the other one ($\Phi_1$) does not couple to them at all; 2) In
contrast, in the type II 2HDM the doublet $\Phi_1$(resp. $\Phi_2$)
couples only to down-like (resp. up-like) quarks. In the latter
case, an additional discrete symmetry in the chiral components of
the fermion sector, namely $f_i \to (-1)^i\,f_i$ (for left- and
right-handed fields $i=1,2$ respectively), must be imposed in order
to banish the tree-level FCNC processes. Let us recall that the MSSM
Higgs sector is a type II one of a very restricted sort (it is
enforced to be SUSY invariant)\,\cite{susy,hunter}. We shall not
dwell here on the Higgs boson interactions with fermions (see
\cite{hunter} for details) because they are not involved in any of
our tree-level Higgs boson production processes (\ref{2H}-\ref{3H}).
Let us finally recall that SUSY invariance of the potential
introduces $5$ constraints that reduce the number of free parameters
to $2$, usually taken to be $M_{\Azero}$ and
$\tan\beta$\,\cite{hunter}. In particular, SUSY decreets that the
two terms softly breaking the discrete symmetry of the potential
must be equal: $\lambda_5 = \lambda_6=2\sqrt{2}\,G_F\,M_{\Azero}^2$.
For simplicity, and in order to keep closer to the MSSM structure of
the Higgs sector, sometimes one adopts this setting \cite{{santi}}.
We will follow this practice and therefore the final number of free
parameters in our analysis will be $6$. They can be arranged as
follows:
\begin{equation}\label{freep}
  (M_{\hzero},M_{\Hzero},M_{\Azero},M_{\Hpm},\tan\alpha,\tan\beta)\,.
\end{equation}

Essential for our calculation are the trilinear Higgs couplings.
These are not explicitly present in Eq.~(\ref{eq:potential}), but
they are generated after spontaneous breaking of the EW symmetry. In
the SM the trilinear and quartic Higgs couplings have fixed values,
which uniquely depend on the actual mass of the Higgs boson. In the
case of the MSSM, and due to the SUSY invariance, the Higgs
self-couplings are of pure gauge nature\,\cite{hunter}. This is in
fact the reason for the tiny triple-Higgs boson production rates
obtained for the processes (\ref{3H}) within the framework of the
MSSM\,\cite{Djouadi:1999gv}, see section \ref{sect:numerical}. In
contrast, the general 2HDM accommodates trilinear Higgs couplings
with great potential enhancement. In Table~\ref{tab:trilinear} we
list those entering our computations, in a form already
accommodating the condition $\lambda_5 = \lambda_6$. These couplings
are written in terms of the physical fields and Goldstone bosons,
which are obtained after rotating the electroweak eigenstates
(\ref{eq:H1H2}) into the physical mass-eigenstates by means of the
two diagonalization angles $\alpha$ and $\beta$. As can be seen, the
couplings in Table~\ref{tab:trilinear}  depend on the $6$ free
parameters (\ref{freep}). The behavior and enhancement capabilities
of these Higgs boson self-interactions are at the heart of our
calculation of the cross-sections (\ref{3H}) within the general 2HDM
(type I and type II).
    \begin{table}[tb]
        \centering
        \begin{tabular}{|c|c|}
            \hline
            $\PHiggspm\PHiggsmp\Hzero$&$-\frac{i\,e}{\mw\sin\theta_W\stbt}
            \left[(\mHps-\mAs+\frac{1}{2}\mHs)\stbt\cbma\right.$\\
            &$\phantom{\frac{g}{\mw\stbt}}\left.-(\mHs-\mAs)\ctbt\sbma\right]$\\\hline
            $\PHiggspm\PHiggsmp\hzero$&$-\frac{i\,e}{\mw\sin\theta_W\stbt}
            \left[(\mHp^2-\mA^2+\frac12\mh^2)\sin{2\beta}\sin(\beta-\alpha)\right.$\\
            &$\phantom{\frac{g}{\mw\stbt}}
            \left.+(\mh^2-\mA^2)\,\cos{2\beta}\,\cos(\beta-\alpha)\right]$\\\hline
            $\hzero\hzero\Hzero$&$-\frac{i\,e\,\cos(\beta-\alpha)}{2\,M_W\sin\theta_W\,
             \sin{2\beta}
             }\,
            \left[(2\,\mh^2+\mH^2)\,\sin{2\alpha}\right.$\\
            &$\phantom{-\frac{g}{\mw\stbt}}
            \left.-\mA^2\,(3\sin{2\alpha}-\sin{2\beta})\right]$\\\hline
            $\hzero\Hzero\Hzero$&$ \frac{i\,e\,\sin(\beta-\alpha)}{2\,M_W \sin\theta_W
             \sin{2\beta}}
            \left[(\mh^2+2\mH^2)\,\sin{2\alpha}\right.$\\
            &$\phantom{-\frac{g}{\mw\stbt}}
            \left.-\mA^2\,(3\sin{2\alpha}-\sin{2\beta})\right]$\\\hline
            $\Azero\Azero\Hzero$&$-\frac{i\,e}{2\,M_{W}\sin\theta_W\sin{2\beta}}\,
            \left[\mH^2\,\sin{2\beta}\cos(\beta-\alpha)\right.$\\
            &$\phantom{-\frac{g}{\mw\stbt}}
            \left.-2(\mH^2-\mA^2)\,\cos{2\beta}\,\sin(\beta-\alpha)\right]$\\\hline
            $\Azero\Azero\hzero$&$-\frac{i\,e}{2\,\mw\sin\theta_W\stbt}
            \left[\mh^2\,\sin{2\beta}\sin(\beta-\alpha)\right.$\\
            &$\phantom{-\frac{g}{\mw\stbt}}
            \left.+2(\mh^2-\mA^2)\,\cos{2\beta}\,\cos(\beta-\alpha)\right]$\\\hline
            $\hzero\hzero\hzero$&$-\frac{3\,i\,e}{\mw\sin\theta_W\stbt}
            \left[\mh^2 [\cos(\beta+\alpha)
            +\frac{1}{2} \sin{2\alpha}\sin(\beta-\alpha)\,] \right.$\\
            &$\phantom{-\frac{g}{\mw\stbt}}
            \left. - M_{A^0}^2 \cos^2(\beta-\alpha) \cos(\beta+\alpha)
            \right]$\\\hline
            $\Hzero\Hzero\Hzero$&$-\frac{3\,i\,e}{\mw\sin\theta_W\stbt}
            \left[M_{H^0}^2 [\sin(\beta+\alpha)
             - \frac{1}{2}\sin2\alpha\cos(\beta-\alpha)] \right.$\\
             &$ \phantom{-\frac{g}{\mw\stbt}}
            \left.- M_{A^0}^2 \sin^2(\beta-\alpha)
             \sin(\beta+\alpha)\right]$\\\hline
            $G^0\Hzero\Azero$&$\frac{\,i\,e}{2\mw\sin\theta_W\stbt}\,\sin(\beta-\alpha)
            \left[\mH^2-\mA^2\right]$
            \\\hline
            $G^0\hzero\Azero$&$-\frac{\,i\,e}{2\mw\sin\theta_W\stbt}\,\cos(\beta-\alpha)
            \left[\mh^2-\mA^2\right]$\\\hline
        \end{tabular}
        \caption{\footnotesize{Trilinear Higgs boson self-interactions in the Feynman
          gauge within the 2HDM. Here $G^0$ is the neutral Goldstone boson. These vertices are involved
          in the Feynman diagrams of Fig.\,\ref{fig:3diagrams} in section
          \ref{sect:numerical}.
          Vertices with gauge bosons are common with the
          MSSM and are not included, see \cite{hunter}.}}
        \label{tab:trilinear}
    \end{table}


However, an important point when studying possible sources of
unconventional physics is to ensure that the SM behavior is
sufficiently well reproduced up to the energies explored so far.
Such a requirement translates into a number of constraints over the
parameter space of the given model. In particular, this severely
limits the a priori enhancement possibilities of the Higgs boson
self-interactions in the 2HDM. First of all we have to keep the
theory within a perturbative regime, which entails that only values
in the approximate range $0.1 \lesssim \tan \beta \lesssim 60$ shall
be allowed. Also very important is to maintain the so-called
(approximate) $SU(2)$ custodial symmetry \cite{custodial}. In models
of physics beyond the SM this is done in terms of the parameter
$\rho$, which defines the ratio of the neutral to charged current
Fermi constants. In general it takes the form
$\rho=\rho_0+\delta\rho$, where $\rho_0$ is the tree-level value. In
any model containing an arbitrary number of doublets (in particular
the 2HDM), we have $\rho_0={M_W^2}/{M_Z^2\cos^2\theta_W}=1$, and
then $\delta\rho$ represents the deviations from $1$ induced by pure
quantum corrections. From the known SM contribution and the
experimental constraints\,\cite{pdg} we must enforce that the
additional quantum effects coming from  2HDM dynamics ought to
satisfy $|\delta\rho_{2HDM}|\le 10^{-3}$. It is thus important to
stay in a region of parameter space where this bound is respected.
Let us recall that the 2HDM one-loop contribution is given
by\,\cite{barbieri83}
\begin{eqnarray}\label{drho}
\delta\rho_{{\rm
2HDM}}&=&\frac{G_F}{8\sqrt{2}\,\pi^2}\left\{M_{H^{\pm}}^2\left[1-\frac{M_{A^0}^2}{M_{H^{\pm}}^2-M_{{A^0}}^2}\,
\ln\frac{M_{H^{\pm}}^2}{M_{A^0}^2}\right]\right.\nonumber\\
&&+\cos^2(\beta-\alpha)\,M_{h^0}^2\left[\frac{M_{A^0}^2}{M_{A^0}^2-M_{h^0}^2}\,
\ln\frac{M_{A^0}^2}{M_{h^0}^2}-\frac{M_{H^{\pm}}^2}{M_{H^{\pm}}^2-M_{h^0}^2}\,
\ln\frac{M_{H^{\pm}}^2}{M_{h^0}^2}\right]\nonumber\\
&&\left.+\sin^2(\beta-\alpha)\,M_{H^0}^2\left[\frac{M_{A^0}^2}{M_{A^0}^2-M_{H^0}^2}\,
\ln\frac{M_{A^0}^2}{M_{H^0}^2}-\frac{M_{H^{\pm}}^2}{M_{H^{\pm}}^2-M_{H^0}^2}\,
\ln\frac{M_{H^{\pm}}^2}{M_{H^0}^2}\right]\right\}\,.
\end{eqnarray}
From this expression it is clear that if $M_{\Azero}\to M_{H^{\pm}}$
then $\delta\rho_{{\rm 2HDM}}\to 0$, and hence if the mass splitting
between $M_{A^0}$ and $M_{H^{\pm}}$ is not significant
$\delta\rho_{{\rm 2HDM}}$ can be kept within bounds.

Also remarkable are the restrictions over the charged Higgs masses
coming from FCNC radiative $B$-meson decays, whose branching ratio
$\mathcal{B}(b \to s \gamma)\simeq 3 \times
10^{-4}$\,\cite{pdg} is measured with sufficient precision to
be sensitive to new physics. The (charged) Higgs boson contribution
to the aforementioned decay is positive and increases when
$M_{\Hpm}$ decreases.  An averaged bound of $M_{\Hpm}
> 350$ $\GeV$ for $\tan \beta \ge 1$ ensues from
\cite{gamba}. It must be recalled that this bound does not apply to
type-I models since for them the charged Higgs couplings to fermions
are proportional to $\cot\beta$ and hence the loop contributions are
highly suppressed at large $\tan\beta$. By the same token too light
charged Higgs boson contributions are also restricted at very low
$\tan\beta\ll 1$ for both type I and type II models.

Apart from these restrictions, and of course respecting the general
Higgs boson mass bounds obtained from unsuccessful searches at LEP
\cite{pdg}, we must consider also the unitarity bounds. Such bounds
come from the fact that the trilinear 2HDM couplings, hereafter
denoted $C(\PHiggsheavy\PHiggsheavy\PHiggsheavy)$, can receive very
large enhancements at high $\tan\beta$. Although unitarity limits
can be presented in a rather detailed and cumbersome way, we shall
avoid cluttering and proceed as in\,\cite{santi}. Therefore, to
assess that the 2HDM remains unitary, we will adhere to the practice
of bounding the size of the 2HDM trilinear Higgs boson couplings by
the value of their single SM counterpart
$\lambda_{\PHiggsheavy\PHiggsheavy\PHiggsheavy}^{(SM)}$ at the scale
of $1~\rm{TeV}$:
\begin{eqnarray}
|C(\PHiggsheavy\PHiggsheavy\PHiggsheavy)|\le \left|
\lambda_{\PHiggsheavy\PHiggsheavy\PHiggsheavy}^{(SM)}(M_{H}=1 \,\TeV) \right|
=\left. \frac{3\,e\,M_{\PHiggsheavy}^2}{2\,\sin\theta_W\,M_W}\right|_{M_{\PHiggsheavy}=1
\,\TeV}\,, \label{eq:unitary}
\end{eqnarray}
where $-e$ is the electron charge and $\theta_W$ is the weak mixing
angle. The range of Higgs boson masses and other 2HDM parameters
ensuing from this condition fall in the ballpark of the more
complete set of (tree-level) unitarity conditions formulated in
various sophisticated -- albeit non fully coincident -- ways in the
literature\,\cite{unitarity}.

\begin{figure}[thb]
\centerline{
\begin{tabular}{c}
\resizebox{!}{1.5cm}{\includegraphics{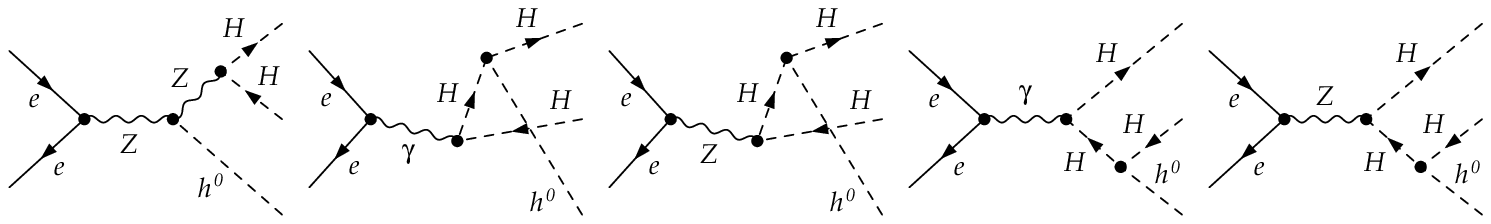}}  \\ \footnotesize{(a)} \vspace{0.2cm} \\
\resizebox{!}{4.75cm}{\includegraphics{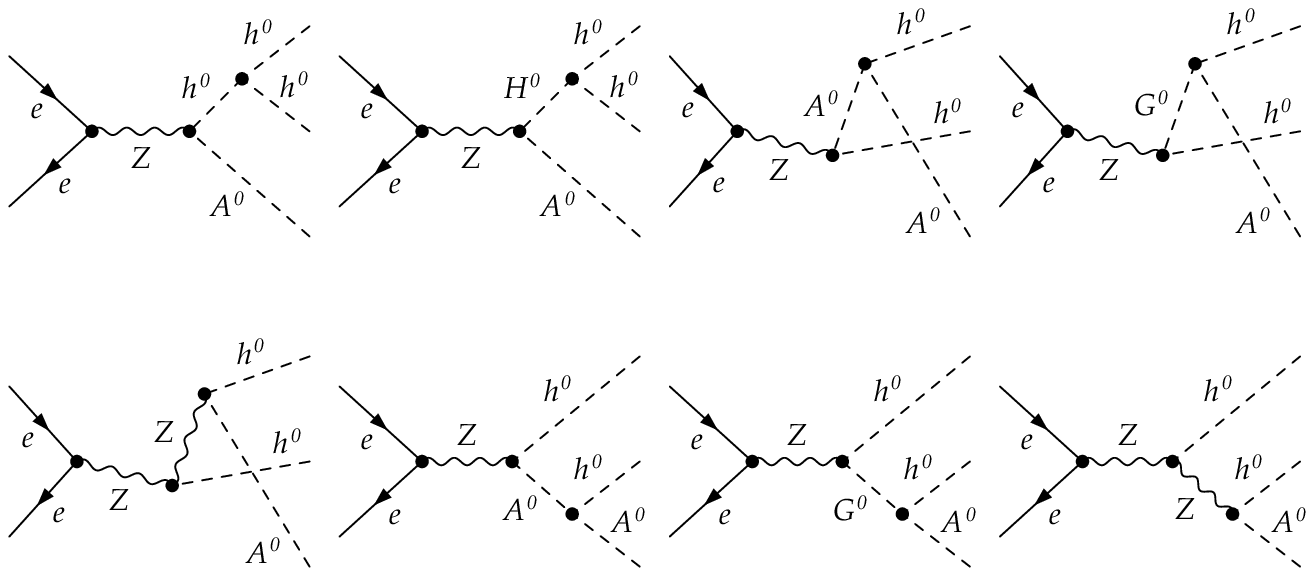}}  \\ \footnotesize{(b)} \vspace{0.2cm}\\
\resizebox{!}{4.75cm}{\includegraphics{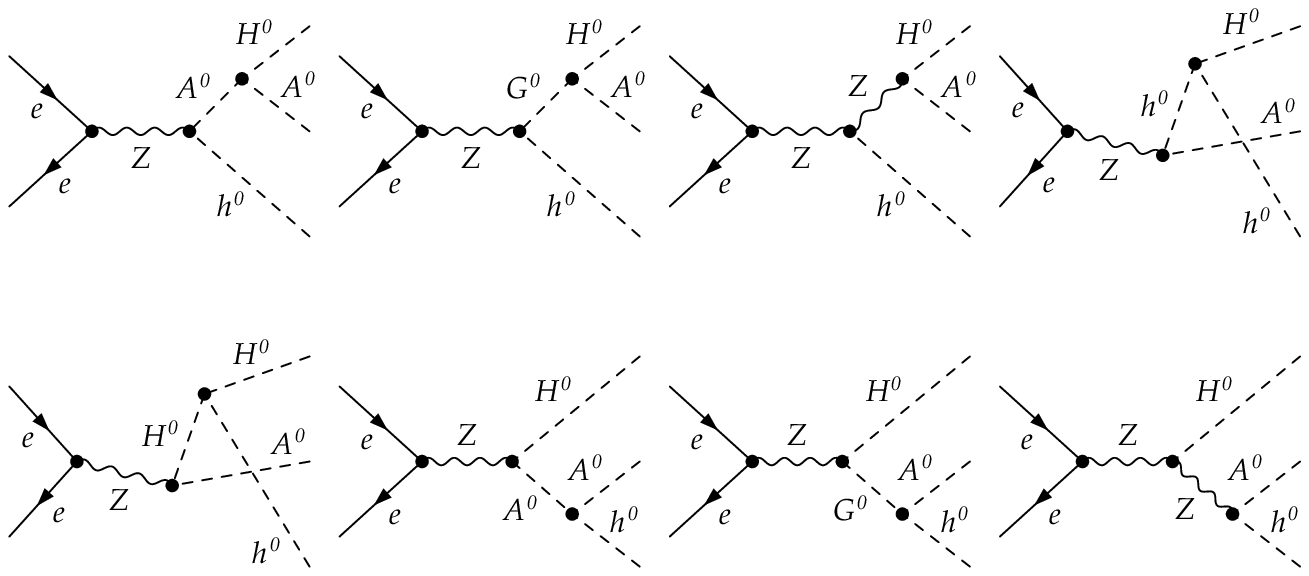}}  \\
\footnotesize{(c)}
 \end{tabular}}
\caption{\footnotesize{Tree-level Feynman diagrams corresponding to
three of the triple Higgs boson production processes indicated in
Eq.\,(\ref{3H}). The other four processes proceed through similar
collections of diagrams.}} \label{fig:3diagrams}
\end{figure}
\begin{figure}
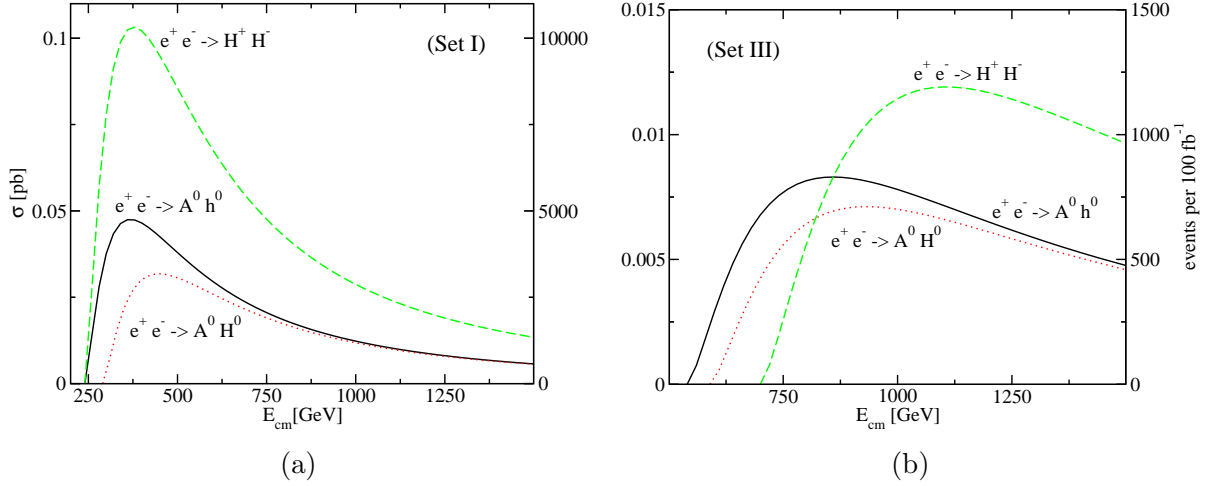

\centerline{
\begin{tabular}{cc}
\resizebox{!}{5.7cm}{\includegraphics[scale=0.35]{twohiggs_light.eps}}
&
\resizebox{!}{5.7cm}{\includegraphics{twohiggs_heavy.eps}} \\
(a) & (b) \\
 \end{tabular}}
\caption{\footnotesize{Total cross section $\sigma$ (in $\picobarn$)
and number of events per $100$ $\invfb$ as a function of ${\rm
E}_{\rm cm}=\sqrt{s}$ for the tree-level Higgs boson pair production
channels (\ref{2H}) in the general 2HDM within two regimes of
masses: \textbf{a)} \textit{light} Higgs bosons and \textbf{b)}
\textit{heavy} Higgs bosons, represented by the mass Sets I and III
respectively in Table \ref{tab:s1}. }} \label{fig:2born}
\end{figure}

\noindent \mysection{Triple Higgs boson production: numerical
analysis} \label{sect:numerical}

Throughout the present work we have made use of the standard
algebraic and numerical packages \textit{FeynArts},
\textit{FormCalc} and \textit{LoopTools} \cite{feynarts} for the
obtention of the Feynman diagrams, the analytical computation and
simplification of the scattering amplitudes and the numerical
evaluation of the cross section.  Feynman diagrams for the
tree-level Higgs-pair production (2H) processes are very simple and
are not shown here, whereas a sample of typical Feynman diagrams for
the triple (3H) Higgs boson production processes is displayed in
Fig.~\ref{fig:3diagrams}. Let us first concentrate on the 2H final
states in the general 2HDM. As indicated in the introduction, these
are well studied in the MSSM case. Here we shall report first on the
tree-level results for 2H production in both the MSSM and general
2HDM, mainly to compare with the triple Higgs boson channels which
are indeed the main aim of the present study.

It is not our intention to present here the one-loop analysis of the
2H processes within the general 2HDM, except to briefly report on
those which can only work at the one-loop level\,\footnote{For
details of the full one-loop analysis, see\,\cite{paper2}.}. For
instance, due to $\CP$-conservation (even more: due to Bose-Einstein
statistics), some of the possible channels (namely
$\HepProcess{\APelectron\,\Pe \HepTo \hzero\,\hzero}$,
$\HepProcess{\APelectron\,\Pe \HepTo \Hzero\,\Hzero}$,
$\HepProcess{\APelectron\,\Pe \HepTo \Azero\,\Azero}$) are forbidden
at the tree level and can take place only through $1$-loop box-type
diagrams. We have evaluated the corresponding cross sections and
turn out to be in the range of $10^{-5}$ $\picobarn$ (that is to
say, they entail around $1$ event only per $100$ $\invfb$ of
integrated luminosity), a too minute rate to provide feasible
detection signals. Notice that in the SM these tree-level forbidden
processes are induced by the same set of box diagrams, and
consequently the SM rates are of the same order. Quite another is
the situation with the other channels, Eq.\, (\ref{2H}), which are
$\mathcal{C}\mathcal{P}-$allowed at the tree-level and hence furnish
sizeable rates. Since only couplings of the guise Higgs-Higgs-gauge
boson play a role, the interactions are of purely gauge nature and
do not differ from the general 2HDM to the MSSM\,\cite{hunter}.
Therefore, we expect both models to provide similar cross-sections
for all the processes (\ref{2H}).
\begin{table}[h]
\begin{center}
\begin{tabular}{|c||c|c|c|} \hline
\quad & Set I & Set II & Set III \\ \hline \hline
$M_{\PHiggslightzero}$\,$(\GeV)$ & 100 & 100 & 200\\ \hline
$M_{\PHiggspm}$\,$(\GeV)$ & 120 & 120 & 350\\\hline
$M_{\PHiggszero}$ \,$(\GeV)$ & 150 & 150 & 250\\\hline
$M_{\PHiggspszero}$\,$(\GeV)$ & 140 & 300 & 340\\ \hline
\end{tabular}
\end{center}
\caption{\footnotesize{Sets I, II and III of \textit{light} and
\textit{heavy} Higgs boson masses in the 2HDM. Sets I and III are
used for 2H production in Fig.\,\ref{fig:2born}, and Sets II and III
for 3H production in Fig.\,\ref{fig:3higgs1} and
\ref{fig:3higgs2}.}} \label{tab:s1}
\end{table}
\begin{table}
\begin{center}
\begin{tabular}{|c|c|c|c|}
\hline \quad & $\sigma_{max}$ ($\sqrt{s}=1$ TeV) & $M_{\Azero}$
($\GeV$) & $\tan\beta$  \\ \hline \hline
$\HepProcess{\APelectron\Pe\HepTo\Azero\hzero}$ &0.013 & 100 & 60
\\ \hline
$\HepProcess{\APelectron\Pe\HepTo\Azero\Hzero}$ &0.012 & 130 & 60 \\
\hline
$\HepProcess{\APelectron\Pe\HepTo\PHiggsplus\PHiggsminus}$& 0.028 &
100 & 5.5 \\ \hline
\end{tabular}
\end{center}
\caption{\footnotesize{Maximum cross sections (in pb) for the 2H
production channels within the MSSM at $\sqrt{s}=1$ TeV.}}
\label{tab:2results}
 \end{table}
\begin{table}[th]
\begin{center}
\begin{tabular}{|c|c|} \hline
$M_{SUSY} (\GeV)$ & 1000 \\ \hline $\mu (\GeV)$ & 200 \\ \hline $A_t
(\GeV)$ & 1000 \\ \hline $A_b (\GeV)$ & 1000 \\ \hline $A_\tau
(\GeV)$ & 1000 \\ \hline
\end{tabular}
\end{center}
\caption{\footnotesize{Choice of parameters used for the computation
of 2H and 3H production in the MSSM.}} \label{tab:mssm}
\end{table}
\begin{figure}[thb]
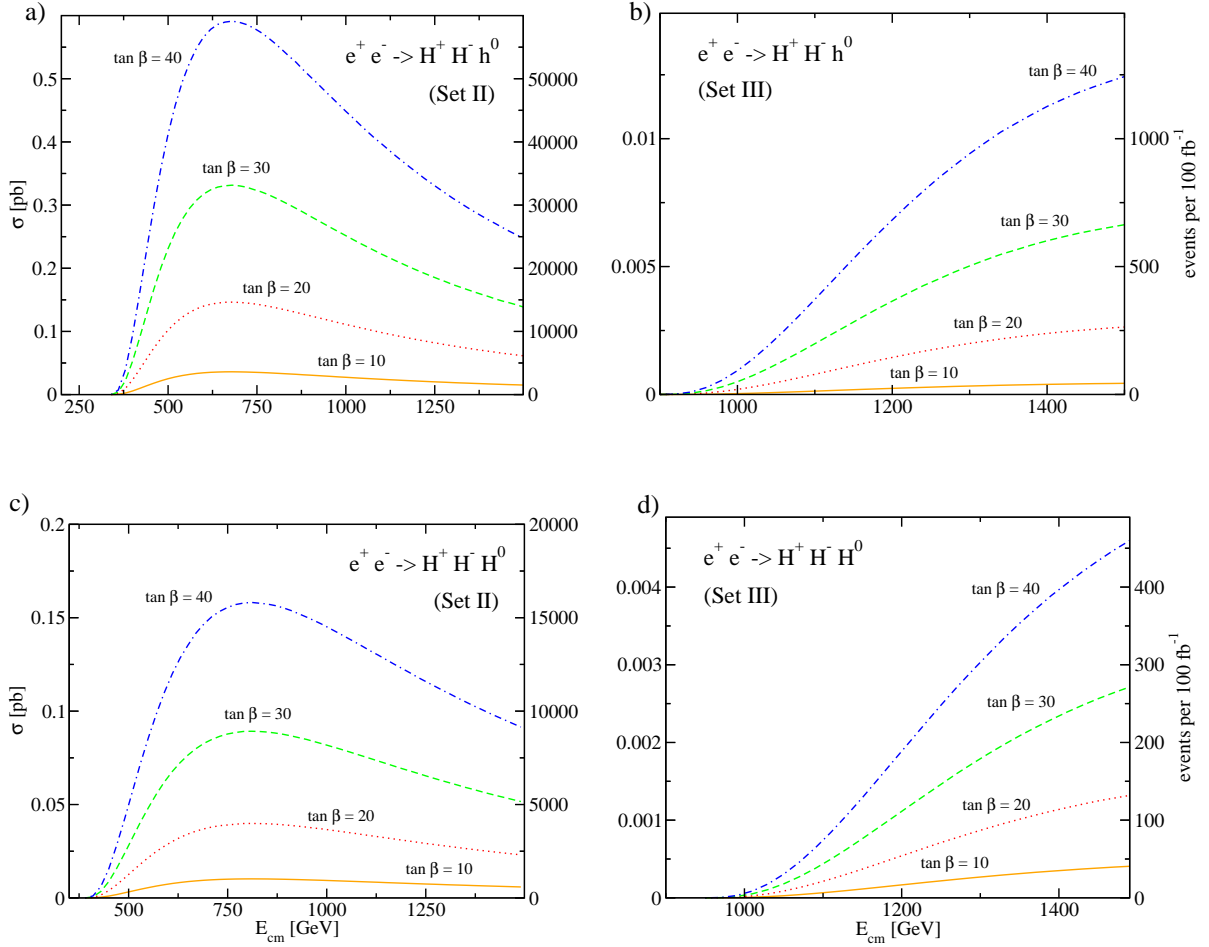

\centerline{
\begin{tabular}{cc}
\resizebox{!}{5.5cm}{\includegraphics{eeHHh_light.eps}} &
\resizebox{!}{5.5cm}{\includegraphics{eeHHh_heavy.eps}}
\\ \vspace{0.3cm} \\
\resizebox{!}{6.cm}{\includegraphics{eeHHH0_light.eps}} &
\resizebox{!}{6.cm}{\includegraphics{eeHHH0_heavy.eps}}
\end{tabular}}
\caption{\footnotesize{Total cross section $\sigma$(pb) and number
of events per $100$ $\invfb$ for the triple Higgs boson production
processes $\eeHHh$ and $\eeHHHzero$ in the general 2HDM as a
function of $\sqrt{s}$ and for different values of $\tan \beta$. In
each case the label of the process and the choice (Set II or Set
III) of Higgs boson masses used for the calculation is indicated,
see Table \ref{tab:s1}. }} \label{fig:3higgs1}
\end{figure}
For the numerical evaluation of the 2H cross sections, we have to
input specific values for the free 2HDM parameters defined in
section \ref{sect:interaction}, see Eq.\,(\ref{freep}). On the one
hand, we will set the mixing angles $\alpha, \beta$ in such a way
that the aforementioned Higgs-Higgs-gauge boson couplings are
optimized. This can be easily done, for instance by setting
$\alpha=\beta$ in $\HepProcess{\APelectron\,\Pe \HepTo
\hzero\,\Azero}$, in which case the relevant coupling
$C(\Azero\hzero\PZzero) \sim \cos(\beta-\alpha)$ is maximum. In
Fig.~\ref{fig:2born} we plot the total cross section $\sigma$ (in
$\picobarn$) as a function of the center-of-mass energy $\sqrt{s}$
(in $\GeV$) for the different channels (\ref{2H}). We explore two
different regimes: \textbf{a)} light Higgs masses and \textbf{b)}
heavy Higgs masses. To represent these regimes we use in this case
sets I and III of Higgs boson masses in Table \ref{tab:s1}. Sets I
and II differ only in the CP-odd Higgs mass, $M_{\Azero}$, which is
substantially lighter in the former as compared to the latter.
Indeed, we wish to use set I for the study of 2H production in order
not to artificially suppress it by mere phase space reasons. On the
other hand, set II (together with set III) will be used later on for
the study of 3H production. From Fig.~\ref{fig:2born}a we see that
in the light Higgs boson mass regime the production rates are
substantial, attaining in all cases a few thousands events per $100$
$\invfb$ of integrated luminosity, with maximum values reaching
$\sigma({\HepProcess{\APelectron\,\Pe \HepTo \PHiggsplus\PHiggsminus}}) \sim 0.1$
$\picobarn$. For heavy Higgs bosons, Fig.~\ref{fig:2born}b tells us
that the achieved production rates are around one order of magnitude
below. Nevertheless, even in these less favored scenarios the
predicted rates are still quite sizeable within the clean ILC
environment.

It is of course of interest to contrast the above 2H results with
the predicted contributions in the MSSM case, see Table
\ref{tab:2results}. Recall from section \ref{sect:interaction} that
the MSSM Higgs sector is fully determined (at the tree-level) by a
pair of free parameters, namely $\tan\beta$ and $M_{\Azero}$. Taking
advantage of this simple structure of the parameter space, we have
systematically searched for the values of ($\tan\beta$,$M_{\Azero}$)
such that the cross section at fixed $\sqrt{s}=1$ $\TeV$ becomes
optimal while simultaneously fulfilling all the phenomenological
constraints on the SUSY Higgs masses. Notice that in the MSSM the
charged Higgs boson mass is not so severely restricted as in the
case of the general 2HDM models of type II (see section
\ref{sect:interaction}) because the squark and chargino
contributions to $\mathcal{B}(b \to s \gamma)$ can compensate for
the charged Higgs effects\,\cite{gamba}. A default set of MSSM
parameters (quoted in Table~\ref{tab:mssm}) has been employed to
compute the Higgs boson masses.  These include the quantum effects
obtained from the standard code provided by the computational tool
package \cite{feynarts}, which implements the results of
Ref.\cite{higgsmass}. We remark that $M_{SUSY}$ in
Table~\ref{tab:mssm} stands for a common value of the LL and RR soft
SUSY-breaking masses in the squark mass matrices. To be sure, the
numerical search for the maximum of $\sigma(2H)$ for the processes
(\ref{2H}) has been made under the condition that $m_{h^0}$ is
larger than its lower experimental bound ($\sim$ 90
$\GeV$\,\cite{pdg}). Let us also note that, in order to get more
accurate results, a running value for the electromagnetic coupling
constant $\alpha(M_Z)= 1/127.9$ has been used. All in all, the
(approximate) optimal values obtained for 2H production within the
MSSM are summarized in Table~\ref{tab:2results}. We see that the
predicted cross sections are of the order of $10^{-2}$ $\picobarn$
and, as anticipated, they are comparable to the 2HDM values for
similar masses. Therefore, sizeable rates of non-standard Higgs-pair
production can be achieved at the ILC for both SUSY and non-SUSY
extended Higgs sectors, and in this sense the two models are
difficult to distinguish using the 2H channels. As mentioned in the
introduction, a clear separation between the two models can only be
accomplished through the detailed study of radiative corrections to
2H production in both the MSSM\,\cite{mssmloop} and the
2HDM\,\cite{paper2}.

Let us now discuss the case of the triple Higgs boson production in
${\ensuremath{\HepProcess{\APelectron\,\Pe^-}}}$ annihilations
within the general 2HDM. The processes under consideration are those
in Eq.\,(\ref{3H}). Again we wish to compute the cross-sections for
them and compare with the corresponding MSSM values. As far as the
2HDM is concerned, we shall keep making use of two separate regimes
of masses, light and heavy, but in this case they will be
represented by sets II and III respectively in Table \ref{tab:s1}.
Actually, due to the low energy $b\to s\,\gamma$ constraint
(mentioned in section \ref{sect:interaction}) on the charged Higgs
boson mass in type II models, we cannot keep the CP-odd mass
$M_{\Azero}$ relatively light (as in set I) for these models. The
distinction between the two sets of masses (sets II and III) is thus
necessary. Set II accommodates higher values of $M_{\Azero}$ than
set I, but only set III reflects a mass region allowed in type-II
2HDM. In fact, due to the constraint $|\delta\rho_{2HDM}|<10^{-3}$
--  cf. Eq.(\ref{drho}) -- it turns out that not only $M_{\Hpm}$,
but also $M_{\Azero}$, must necessarily be heavier in type II
models.

\begin{figure}[bht]
\centerline{
\begin{tabular}{cc}
\resizebox{!}{5.5cm}{\includegraphics{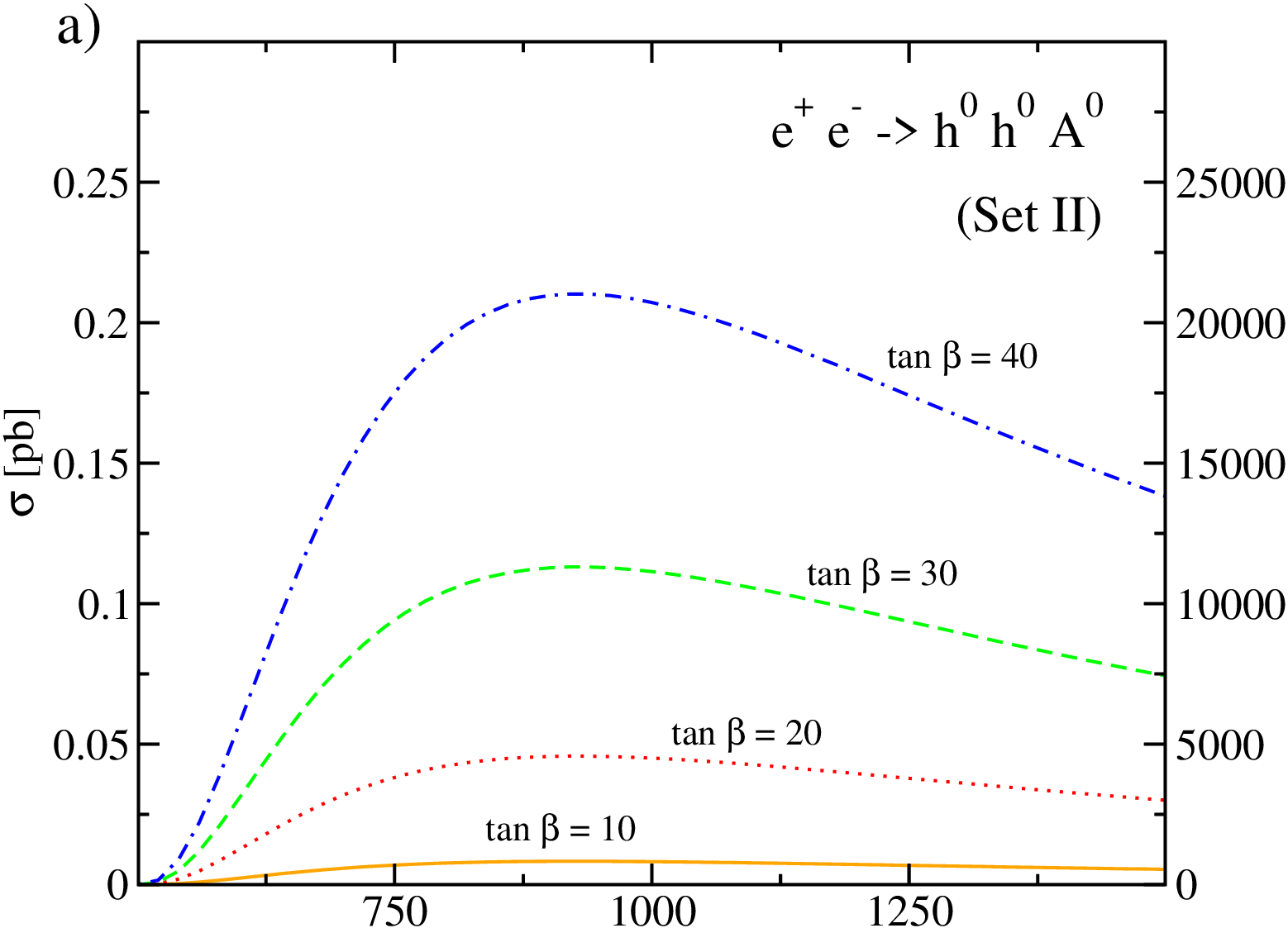}} & \quad
\resizebox{!}{5.5cm}{\includegraphics{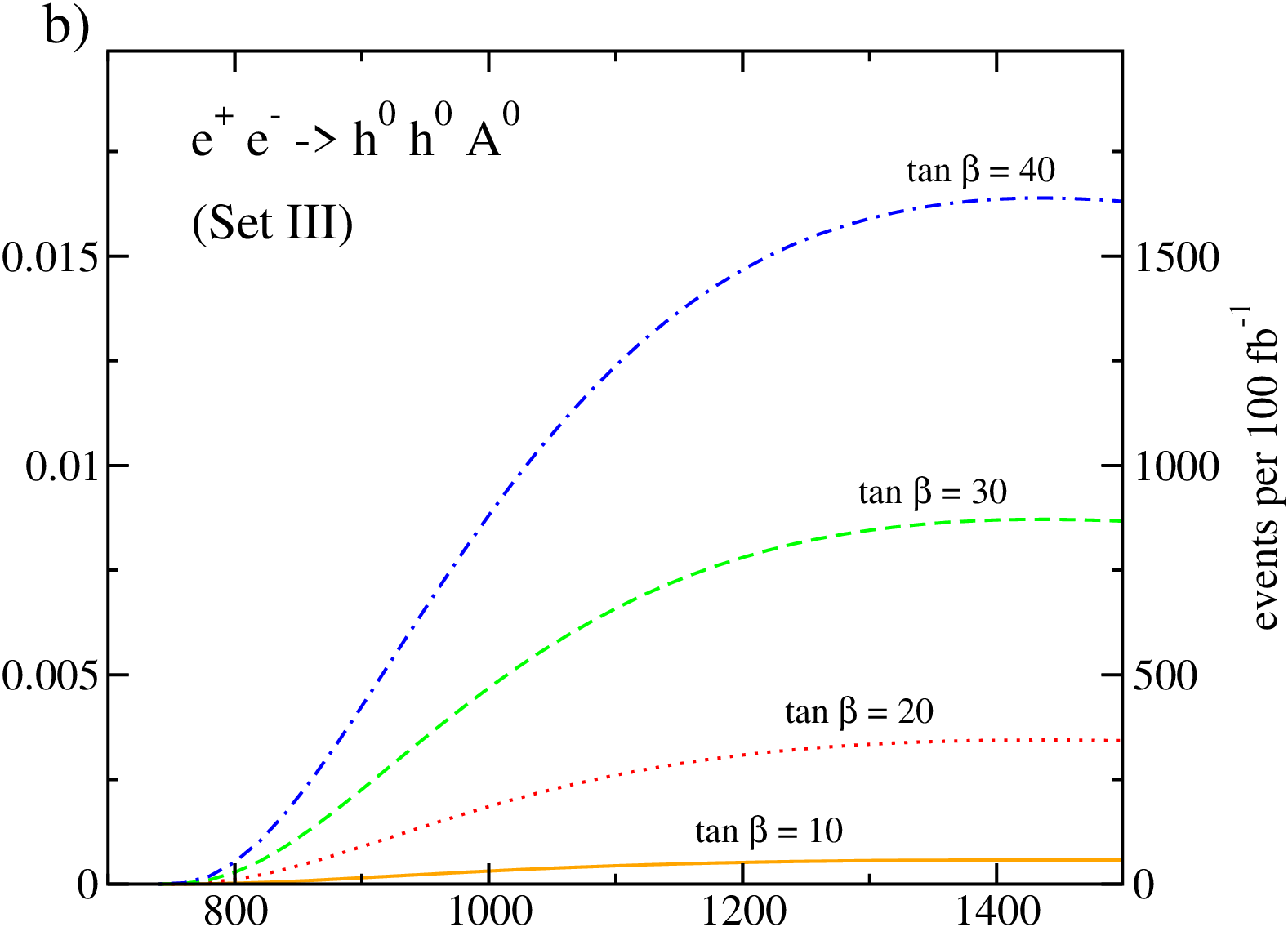}} \\ \vspace{0.3cm} \\
\resizebox{!}{5.9cm}{\includegraphics{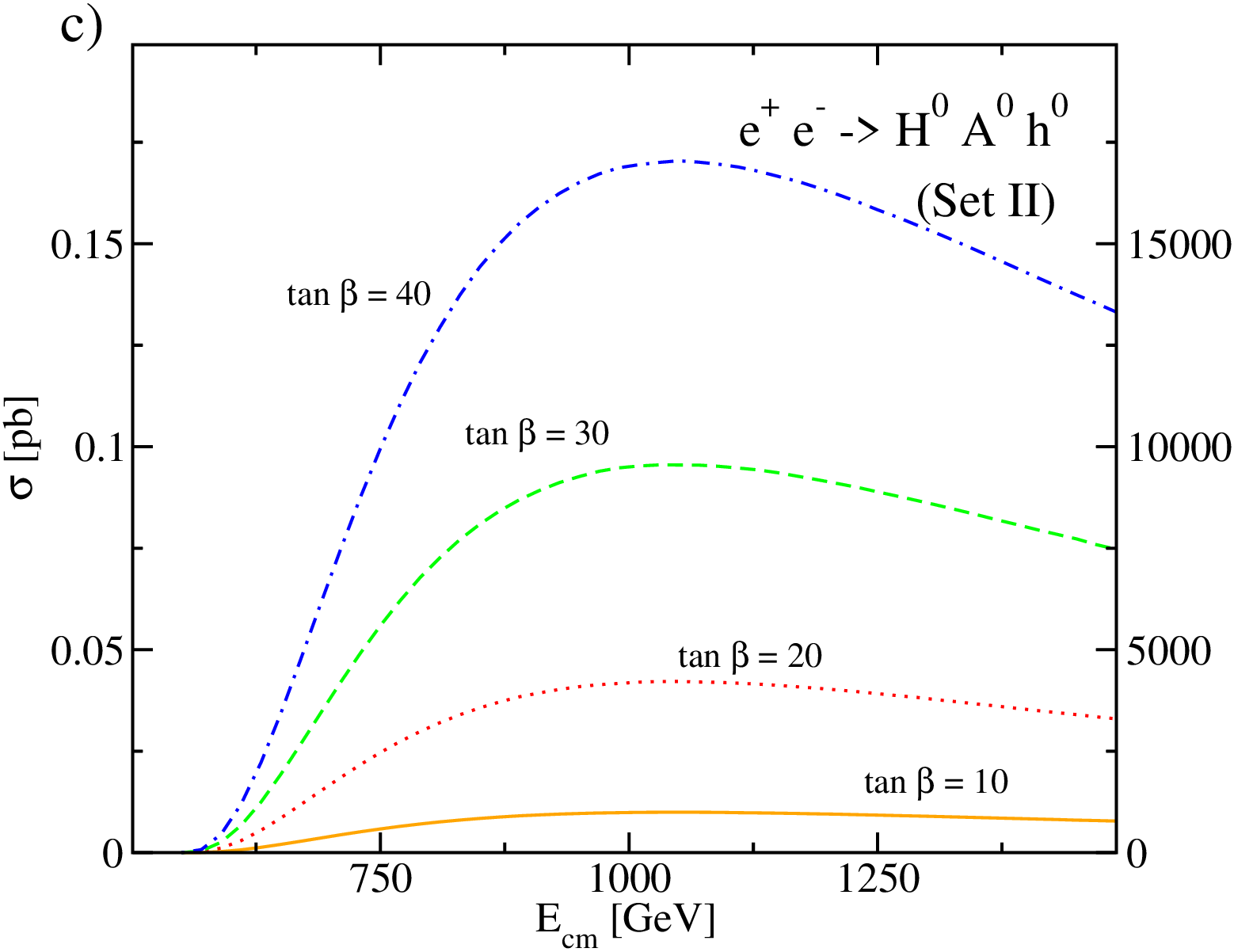}} & \quad
\resizebox{!}{5.9cm}{\includegraphics{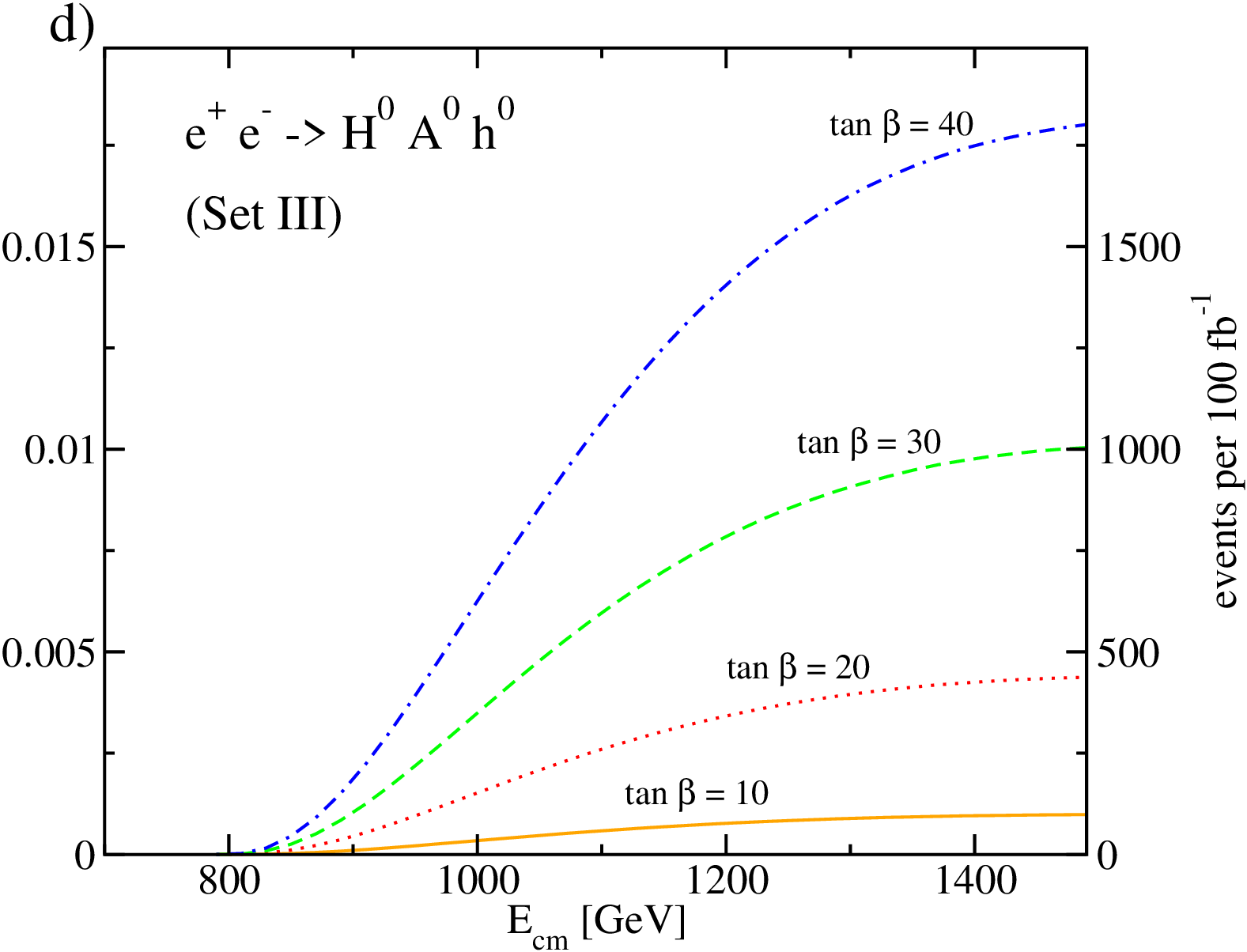}}
 \end{tabular}}
\caption{\footnotesize{As in Fig.\,\ref{fig:3higgs1}, but for
processes $\eehhA$ and $\eeHzeroAh$.} \vspace{0.5cm}}
\label{fig:3higgs2}
\end{figure}

The Feynman diagrams for the most prominent triple Higgs boson
processes (\ref{3H}) are depicted in Fig.\,\ref{fig:3diagrams}. We
can see that they involve trilinear Higgs boson couplings of the
form indicated in Table \ref{tab:trilinear}. Let us illustrate the
origin of the potential enhancement inherent to these couplings by
just focusing on one of them, for example the first one, $C(\Hpm
\,\Hpm\,\Hzero)$. One can easily check that for $\tan\beta \gg 1$ or
$\tan\beta \ll 1$ and $\alpha\simeq 0$, the coupling grows
effectively as $\sim\tan\beta$ or $\sim\cot\beta$ respectively.
Therefore, the corresponding cross section can be significantly
enhanced by a factor $\tan^2\beta$ or $\cot^2\beta$ respectively. We
will mainly explore the enhancement in the large $\tan\beta$ region,
which is more natural and also more efficient. An additional
enhancement source in $C(\Hpm \,\Hpm\,\Hzero)$ is the possible mass
splittings between the Higgs boson masses, e.g. between $\mA^2$ and
$\mH^2$, which is also subdued in part by the $\delta\rho_{\rm
2HDM}$ constraint mentioned above. In contrast to this situation, in
the MSSM the triple Higgs couplings do not have any such
enhancements. Indeed, in the MSSM the (tree-level) analogous of the
coupling under consideration is
\begin{eqnarray}
C_{\mbox{\itshape\tiny {\rm MSSM}}}(\Hpm \,\Hpm\,\Hzero)&=&
 \frac{-i e M_W}{\sin\theta_W}\left[\cos{(\beta-\alpha)-\frac{\cos{2\beta}\,\cos{(\alpha+\beta)}}{2\cos^2\theta_W}
}\right].
\end{eqnarray}
It is patent that there is no source of enhancement, the coupling
being gauge-like. In general the Higgs boson self-couplings in the
MSSM undergo radiative corrections \,\cite{3HcorrMSSM} (as the Higgs
boson masses themselves), but in practice the 3H cross sections
remain rather small\,\cite{pairmssm,Djouadi:1999gv,Fawzy02}. In
contrast, the enhancement effect of the general 2HDM trilinear
couplings, listed in Table \ref{tab:trilinear}, can have a much
bigger impact on the cross-sections while respecting all known
bounds. In this respect, we recall that these couplings can also
receive radiative corrections in the general
2HDM\,\cite{3Hcorr2HDM}. However, we do not include them here
because our main goal is to show that the 3H production signal can
be significantly enhanced in the general 2HDM, and for this it
suffices to confine ourselves to the tree-level structure. More
refined studies of this signal may be necessary in the future, in
which case the inclusion of corrections should be appropriate.

We have plotted the 3H cross-sections for the general 2HDM in
Fig.\,\ref{fig:3higgs1} and \ref{fig:3higgs2} for the Higgs boson
mass sets II and III in Table \ref{tab:s1}. We see that they can
reach the level of $\sim 0.1$ pb or more, therefore implying
promising rates of at least $10^4$ events per $100~\rm{fb}^{-1}$ of
integrated luminosity. The larger cross sections are obtained
considering light Higgs masses (set II of Tab.~\ref{tab:s1}). This
is because there is a lower suppression of the final phase-space and
also because the maximum of the cross section is reached at lower
energies $\sqrt{s}\sim (700- 1000)~\rm{GeV}$. Furthermore, as
expected, all the cross-sections are seen to increase approximately
as $\tan^2\beta$ due to the behavior of the trilinear couplings. In
the heavy Higgs boson scenario (set III), the maximum is shifted to
higher values $\sqrt{s}  \sim 1500$~GeV. Taking into account the
presence of the $Z$ boson propagator, the cross section scales with
the energy and thus the corresponding maximum becomes between one to
two orders of magnitude smaller than in the light Higgs boson
scenario. These results for set III (adequate to type II models)
translate into rates of ${\cal O}(10^2-10^3)$ events per
$100~\rm{fb}^{-1}$ of integrated luminosity, which should still
allow comfortable detection of the signal. As for the remaining
Higgs production channels (not considered in our figures), they
provide smaller values of the maximum cross-section: e.g. those with
final states $\PHiggsplus\PHiggsminus\Azero$ and $\Hzero \Hzero
\Azero$ yield maximum cross-sections of order $10^{-2}$ pb for set
II and $(10^{-3}-10^{-4})$ pb for set III. Finally, channel $\Azero
\Azero \Azero$ is a rather inconspicuous one due to the phase-space
suppression and also to the fact that we have three identical
particles in the final state (hence an additional suppression of
$1/3!$), leaving maximum cross-sections of order $10^{-4}\, \rm{pb}$
and $10^{-5}\, \rm{pb}$ for sets II and III at
$\sqrt{s}=1400~\rm{GeV}$.

Some technical details are now in order. To find the values of the
angles $\alpha$ and $\beta$ that generate the maximum of the
cross-section $\sigma_{max}(3H)$ for the various 3H processes we
have performed a systematic scan using the sets II and III of
parameters given in Table~\ref{tab:s1} under the restrictions
mentioned in section \ref{sect:interaction}. The result is that in
all cases the largest possible values of $\tan\beta$ are preferred
rather than intermediate or very small ones. However, $\tan\beta$
cannot be arbitrarily large (or arbitrarily small), not only due to
the perturbative bound, but also because of the unitarity
constraint. For this reason in Figs.~\ref{fig:3higgs1} and
\ref{fig:3higgs2} we have limited ourselves to plot the
cross-sections for a few values of $\tan\beta$ up to $\tan\beta=40$.
We have already exemplified how some trilinear couplings are
maximized e.g. for large $\tan\beta$ and $\alpha\simeq 0$, but
others are not so enhanced in this region. Our numerical scan shows
that the following intermediate strategy optimizes the
cross-sections: once $\tan\beta$ is chosen at the largest allowed
value, we choose $\alpha=\pi/2-\beta-\delta$, typically with
$\delta=0.8$ (rad.). This is enough to circumvent the unitarity and
perturbative restriction and get the optimal set of cross-sections.
These are the ones studied in Figs.~\ref{fig:3higgs1} and
\ref{fig:3higgs2}. Let us also point out that the maximum in each
process is not severely peaked; we have verified that there is a
large region of parameter space (including $\alpha\simeq\beta$)
where the cross-sections are still perfectly sizeable (within the
same order of magnitude as $\sigma_{max}$). For completeness, let us
also mention that the corresponding values of the parameter
$\lambda_5$ in Figs.\,\ref{fig:3higgs1} and \,\ref{fig:3higgs2} are
around $\lambda_5=3$ for set II and $\lambda_5=4$ for set III.
\begin{table}[tb]
\begin{center}
\begin{tabular}{|c|c|c|c|c|}
\hline \quad & $\sigma_{max}$ (1 $\TeV$) & $\sigma_{max}$ (1.4
$\TeV$) & $M_{\Azero}$ ($\GeV$) &
$\tan\beta$  \\ \hline \hline
$\eeHHh$ & $5.6\times10^{-6}$ & $3.6\times10^{-6}$ & 135 & 3 \\
\hline
$\eeHHHzero$ & $1.5\times10^{-6}$ & $9.1\times10^{-7}$ & 100 & 30
\\ \hline
$\eehhA$ & $1.2\times 10^{-3}$ & $7.3\times 10^{-4}$ & 200 & 2.5 \\
\hline
$\eeHzeroAh$ & $2.0\times10^{-6}$ & $1.4\times 10^{-6}$ &100 & 5.5 \\
\hline
\end{tabular}
\end{center}
\caption{\footnotesize{Maximum cross-sections (in pb) for the
leading 3H processes within the MSSM at two values of the center of
mass energy, $\sqrt{s}=1$ TeV and $1.4$ TeV. The maximizing values
of $M_{\Azero}$ and $\tan\beta$} are also indicated and are
(approximately) the same at the two energies. The 3H processes
non-included are even more suppressed.} \label{tab:3mssm}
 \end{table}

To compare the 2HDM results with the corresponding supersymmetric
values we have computed all the 3H production rates $\sigma(3H)$ in
the framework of the MSSM. We have searched for the optimal regions
of the MSSM parameter space where the largest allowed values for the
cross sections are obtained. Specifically, in Table~\ref{tab:3mssm}
we provide the maximum value that $\sigma(3H)$ can achieve for each
process and for two different values of the center-of-mass energy
($\sqrt{s}= 1$ $\TeV$ and $1.4$ TeV) after scanning on
$(M_{A^0},\tan\beta)$ for the fixed values of the MSSM parameter
space quoted in Table~\ref{tab:mssm}. Again the latter determine the
Higgs boson masses at the quantum level from the results of
Ref.\cite{higgsmass}. Let us notice from Table~\ref{tab:3mssm} that
the channel $\eehhA$ has a cross-section that is substantially
larger than the others, the reason being that it can pick up the
resonant decay $\Hzero\to\hzero\,\hzero$ whose branching ratio is
non-negligible in these conditions\,\cite{pairmssm}. As a
consequence $\sigma(\eehhA)$ is of the order of an average 2H
cross-section (cf. Table \ref{tab:2results}) times this branching
ratio. This effect has been studied in detail by including the MSSM
radiative corrections to the trilinear coupling, which turn out to
be important in this region and are responsible for ${\cal
B}(\Hzero\to\hzero\,\hzero)$ being sizeable (of order $50\%$). As a
result the cross-section can be ${\cal O}(10^{-3})$ pb, i.e. of a
few fb. This situation is special in the MSSM, and an accurate
evaluation of it depends on the specific choice of parameter values,
see \,\cite{pairmssm,Djouadi:1999gv}. The great enhancement
associated with it gives some hope for measuring this particular 3H
channel in the MSSM.

In the general 2HDM, that resonant situation is not especially
noticeable because the 3H channels are usually of the same order as
the 2H ones, if not dominant. Therefore, barring that resonant
process, in all the other cases the MSSM cross sections for 3H
production are very small, reaching maximum values of $\sigma \sim
10^{-6}$ pb at most for the leading processes indicated in Table
\ref{tab:3mssm}. The remaining 3H channels in the MSSM, namely those
with final states ${\rm H}^+ {\rm H}^-\Azero$, $\Hzero \Hzero
\Azero$ and $\Azero \Azero \Azero$, furnish maximum cross-sections
in the range $(10^{-7}-10^{-8})$ pb. As a matter of fact, we can
assert that most of the 3H cross-sections in the MSSM are of the
same order as -- if not smaller than -- the tiny rates for the
one-loop 2H processes $\APelectron \Pelectron \to \PHiggslight
\PHiggslight$ (with two identical Higgs particles in the final
state) mentioned in the beginning of this section. In short, we
conclude that the maximum MSSM cross-sections for 3H production are
typically $10^4$ times smaller than the corresponding maximum 2HDM
values (even if taking set III of Higgs boson masses). In the light
of these results it becomes clear that the triple Higgs boson
channels are in general much more promising in the 2HDM (both in
type I and II) than in the MSSM, and can be fully competitive with
the 2H ones.

\noindent \mysection{Discussion and conclusions}
\label{sect:conclusions}

We have devoted this work to the study of the triple Higgs boson
final states (\ref{3H}) produced in a linear $\APelectron\Pelectron$
collider. We have computed the cross-sections for these processes
both in the Minimal Supersymmetric Standard Model (MSSM) and in the
general Two-Higgs-Doublet Model (2HDM).  The results are in
principle independent of which kind of 2HDM model is used, type I or
type II, because the 3H processes (\ref{3H}) are not sensitive to
the Higgs boson interactions with fermions. However, radiative
B-meson decays (characterized by the $b\rightarrow s\gamma$
subprocess) place an important constraint on the lower value of the
charged Higgs boson mass of type II models, namely
$M_{H^{\pm}}\gtrsim 350~\rm{GeV}$, and this fact is what actually
puts an upper bound to the 3H cross-sections for type II models. We
have found that within the type I model the triple Higgs boson
cross-sections may comfortably reach $0.1$pb for $\tan\beta$
sufficiently large ($\tan\beta\gtrsim 20$) or small
($\tan\beta<0.1$)\,\footnote{The neighborhood $\tan\beta\lesssim
0.1$ borders the perturbativity limit of the top quark Yukawa
coupling, and thus the region $\tan\beta<1$ becomes rapidly
excluded.}; actually, in certain regions of parameter space they can
be pushed up to $1$pb, the most favorable process being $\eeHHh$.
This is also the preferred channel for type II models, but due to
the aforesaid charged Higgs boson mass bound the maximum
cross-section is roughly $10$ times smaller, i.e. of order of
$0.01$pb. The number of events is nonetheless of order $10^3$ per
$100\,$fb$^{-1}$ of integrated luminosity, and in both cases the
cross-section is far larger than in the MSSM. For example, the
maximum cross-section for $\eeHHh$ in the MSSM is at most of order
$10^{-6}$pb, i.e. around $10^4$ times smaller than the corresponding
one in general type II Higgs boson models (of which the MSSM Higgs
sector is a very particular case).

Another remarkable fact that we would like to emphasize is that for
the general 2HDM models the maximum cross-sections for the 3H
processes (\ref{3H}) are comparable or even larger than the maximum
cross-sections for the 2H processes (\ref{2H}). Notice that, in
spite of having one more particle in the final state, the mechanism
of 3H production is peculiar in that it involves certain trilinear
Higgs boson couplings that can be enhanced in the general 2HDM, e.g.
at large $\tan\beta>20$. This enhancement is impossible in the MSSM,
due to the purely gauge nature of the Higgs boson self-interactions
in this model which is enforced by the invariance of the potential
under supersymmetric transformations. Incidentally, the 2H
cross-sections for the unconstrained 2HDM models are of the same
order as the 2H cross-sections in the MSSM. In view of these facts,
we expect that the 3H production channels in the general 2HDM could
be competitive at the ILC and provide a direct window for uncovering
the structure of the Higgs potential.

We have found that the regions of parameter space with the largest
possible values of $\tan\beta$ and relatively small $\alpha$ turn
out to maximize the 3H cross-sections. For type II models
(characterized by a heavier spectrum of Higgs boson masses) this
means that the dominant decay modes for each of the Higgs bosons in
a typical final state like $\PHiggsplus\PHiggsminus\hzero$ will be
into heavy quarks. Specifically, the neutral Higgs boson will decay
as $\hzero\rightarrow \Pbottom \APbottom$ and the charged ones as
$\PHiggsplus\rightarrow \Ptop \APbottom$ and
$\PHiggsminus\rightarrow \APtop \Pbottom$. The last two decays
assume of course $M_{H^{\pm}}>m_t+m_b$, which is indeed always the
case due to the $\Pbottom\rightarrow \Pstrange \Pphoton$ constraint
for type II models. In this region of parameter space, the alternate
Higgs boson decays into gauge bosons (such as $\hzero\rightarrow
\PWplus\PWminus,\PZ\PZ$) are not dominant -- in contradistinction to
the SM Higgs boson decays. Other modes like $\Hpm\to \PWpm \hzero$,
even if kinematically open, are suppressed by trigonometric factors
in the coupling strength, viz. $\cos(\beta-\alpha)\to 0$ in the
favorable regions for 3H production. In this region, typically $2/3$
of the Higgs boson decays contribute to the $6$ heavy-quark jet
final states, the rate being larger the larger is $\tan\beta$. In
practice we would expect seeing a $4$-prong final state made out of
$\Pbottom$- and $\APbottom$-jets together with a $\Ptop\APtop$
system decaying in the conventional manner. This configuration
represents the characteristic signature of the 3H processes under
consideration. Although a dedicated experimental study would be
necessary to assess its real possibilities, we expect that in the
extremely clean context of the ILC this signature could hardly be
missed and, if effectively found, it would represent a strong hint
of (non-supersymmetric) Higgs boson physics beyond the SM.

\vspace{0.2cm}
 \noindent
\textbf{Acknowledgments}. This work has been supported in part by
the EU project RTN MRTN-CT-2006-035505 Heptools; GF thanks an ESR
position of this network, and the hospitality of the Dept. ECM,
Univ. de Barcelona where this work was carried out. DLV has been
supported by the MEC FPU grant AP2006-00357; JG and JS are also
supported by MEC and FEDER under project 2004-04582-C02-01 and by
DURSI Generalitat de Catalunya under project 2005SGR00564.
\newcommand{\JHEP}[3]{ {JHEP} {#1} (#2)  {#3}}
\newcommand{\NPB}[3]{{\sl Nucl. Phys. } {\bf B#1} (#2)  {#3}}
\newcommand{\NPPS}[3]{{\sl Nucl. Phys. Proc. Supp. } {\bf #1} (#2)  {#3}}
\newcommand{\PRD}[3]{{\sl Phys. Rev. } {\bf D#1} (#2)   {#3}}
\newcommand{\PLB}[3]{{\sl Phys. Lett. } {\bf B#1} (#2)  {#3}}
\newcommand{\EPJ}[3]{{\sl Eur. Phys. J } {\bf C#1} (#2)  {#3}}
\newcommand{\PR}[3]{{\sl Phys. Rep } {\bf #1} (#2)  {#3}}
\newcommand{\RMP}[3]{{\sl Rev. Mod. Phys. } {\bf #1} (#2)  {#3}}
\newcommand{\IJMP}[3]{{\sl Int. J. of Mod. Phys. } {\bf #1} (#2)  {#3}}
\newcommand{\PRL}[3]{{\sl Phys. Rev. Lett. } {\bf #1} (#2) {#3}}
\newcommand{\ZFP}[3]{{\sl Zeitsch. f. Physik } {\bf C#1} (#2)  {#3}}
\newcommand{\MPLA}[3]{{\sl Mod. Phys. Lett. } {\bf A#1} (#2) {#3}}


\begin{thebibliography}{99}
\bibitem{susy} H.P Nilles, \PR{110}{1984}{1}; H.E. Haber and G.L. Kane,
\PR{117}{1985}{75}.
%
\bibitem{hunter}J.F. Gunion, H.E. Haber, G.L. Kane and S. Dawson,
\textit{The Higgs hunter's guide}, Addison-Wesley, Menlo-Park, 1990.
%
\bibitem{Djouadi:1992pu} A. Djouadi, H.E. Haber and P.M. Zerwas,
\ZFP{57}{1993}{569}.
%
\bibitem{pairmssm} A. Djouadi, H.E. Haber and P.M. Zerwas, \PLB{375}{1996}{203},
 \texttt{hep-ph/9602234}; A. Djouadi, V. Driesen, W. Hollik and J. Rosiek,
 \NPB{491}{1997}{68}, \texttt{hep-ph/9609420}.
%
\bibitem{Feng:1996xv}  J.L. Feng and T. Moroi, \PRD{56}{1997}{5962}, \texttt{hep-ph/9612333}
%
\bibitem{mssmloop} V. Driesen, W. Hollik and J. Rosiek, (1996);
E. Coniavitis and A. Ferrari, \PRD{75}{2007}{015004}; S. Heinemeyer,
\IJMP{21}{2006}{2659}.
%
\bibitem{Djouadi:1999gv} A. Djouadi, W. Kilian, M. Muhlleitner and P.M. Zerwas,
\EPJ{10}{1999}{27}, \texttt{hep-ph/9903229}. For tree-level double
Higgs production processes, see e.g. the exhaustive overview by M.
Muhlleitner, \texttt{hep-ph/0008127}.
%
\bibitem{Fawzy02} P. Osland and P.N. Pandita, \PRD {59} {1998}{055013}; D.J. Miller and S. Moretti,
\EPJ {13} {2000} {459}; F. Boudjema and A. Semenov, \PRD {66}
{2002}{095007}.
%
\bibitem{Weiglein:2004hn} G.~Weiglein et~al. Phys. Rept. \textbf{426}, 47-358 (2006),
\texttt{hep-ph/0410364}.
%
\bibitem{ghk} J. Guasch, W. Hollik, A. Kraft, \NPB {596} {2001} {66}.
%
\bibitem{inlhc} A. Djouadi, W. Kilian, M. Muhlleitner and P.M. Zerwas,
\EPJ{10}{1999}{45}, \texttt{hep-ph/9904287};
T. Binoth, S. Karg, N. Kauer and R. Ruckl, \PRD{74}{2006}{113008}, \texttt{hep-ph/0608057}.
%
\bibitem{paper2} G. Ferrera, J. Guasch, D. L\'opez-Val and J. Sol\`a, work in progress.
%
\bibitem{santi} S. B\'ejar, J. Guasch and J. Sol\`a, \NPB{600}{2001}{21}, \texttt{hep-ph/001091};
S. B\'ejar, J. Guasch and J. Sol\`a, \NPB{675}{2003}{270},
\texttt{hep-ph/0307144}; S. B\'ejar, (2006), PhD Thesis,
\texttt{0606138}.
%
\bibitem{custodial} M.B. Einhorn, D. R. T. Jones and M. J. G. Veltman,
\NPB{191}{1981}{146}.
%
\bibitem{pdg} W. M. Yao et al. (Particle Data Group Collaboration), \textit{J. Phys.} \textbf{G33} (2006) 1.
%
\bibitem{barbieri83} R. Barbieri and L. Maiani, \NPB{224}{1983}{32}.
%
\bibitem{gamba} P. Gambino and M. Misiak, \NPB{611}{2001}{338}, \texttt{hep-ph/0104034}; M. Ciuchini, G. Degrassi, P. Gambino, and G. F. Giudice, \NPB{527}{1998}{21}, \texttt{hep-ph/9710335}; F. M. Borzumati and C. Greub, \PRD{58}{1998}{074004}, \texttt{hep-ph/9802391};
F. M. Borzumati and C. Greub, \PRD{59}{1999}{057501},
\texttt{hep-ph/9809438} (Addendum).
%
\bibitem{unitarity} S. Kanemura, T. Kubota and E. Takasugi, \PLB{313}{1993}{155}, \texttt{hep-ph/9303263};
A. G. Akeroyd, A. Arhrib and E.-M. Naimi, \PLB{490}{2000}{119}, \texttt{hep-ph/0006035};
J. Horejsi and M. Kladiva, \EPJ{46}{2006}{81}, \texttt{hep-ph/0510154}.
%
\bibitem{feynarts} T. Hahn, \textit{FeynArts 3.2, FormCalc}
and \textit{LoopTools} user's guides, available from
\texttt{http://www.feynarts.de}; T. Hahn, \textit{Comput. Phys.
Commun.} \textbf{168} (2005) 78, \texttt{hep-ph/0404043}.

%
\bibitem{higgsmass} S. Heinemeyer, W. Hollik and G. Weiglein, \PLB{455}{1999}{179},
\texttt{hep-ph/9903404}.

\bibitem{3HcorrMSSM}  W. Hollik, S. Pe\~naranda,  \EPJ {23} {2002} {163}, \texttt{hep-ph/0108245};
A. Dobado, M.J. Herrero, W. Hollik and S. Pe\~naranda,  \PRD
{66}{2002} {095016}, \texttt{hep-ph/0208014};
\texttt{hep-ph/0210315}.

\bibitem{3Hcorr2HDM} S. Kanemura, Y. Okada, E. Senaha, \PLB
{606}{2005}{361}, \texttt{hep-ph/0411354}; \texttt{hep-ph/0507259}.

\end{thebibliography}
\end{document}